\documentclass[journal=jctc,manuscript=article]{achemso}
\usepackage[version=3]{mhchem} 
\usepackage{graphicx, color, float, array,amsfonts,hyperref}

\newcommand*{\citen}[1]{%
  \begingroup
    \romannumeral-`\x 
    \setcitestyle{numbers}%
    \cite{#1}%
  \endgroup   
}

\author{Dhiman Ray}
\email{dhiman.ray@iit.it}
\author{Michele Parrinello}
\email{michele.parrinello@iit.it}
\affiliation[]
{Atomistic Simulations, Italian Institute of Technology, Genoa, Via Enrico Melen 83, GE 16153, Italy }

\title[]
{Data Driven Classification of Ligand Unbinding Pathways}
\abbreviations{IR,NMR,UV}
\keywords{American Chemical Society, \LaTeX}

\begin{document}


\newpage
\begin{abstract}
\noindent
Studying the pathways of ligand-receptor binding is essential to understand the mechanism of target recognition by small molecules. The binding free energy and kinetics of protein-ligand complexes can be computed using molecular dynamics (MD) simulations, often in quantitative agreement with experiments. However, only a qualitative picture of the ligand binding/unbinding paths can be obtained through a conventional analysis of the MD trajectories. Besides, the higher degree of manual effort involved in analyzing pathways limits its applicability in large-scale drug discovery. Here we address this limitation by introducing an automated approach for analyzing molecular transition paths with a particular focus on protein-ligand dissociation. Our method is based on the dynamic time-warping (DTW) algorithm, originally designed for speech recognition. We accurately classified molecular trajectories using a very generic descriptor set of contacts or distances. Our approach outperforms manual classification by distinguishing between parallel dissociation channels, within the pathways identified by visual inspection. Most notably, we could compute exit-path-specific ligand-dissociation kinetics. The unbinding timescale along the fastest path agrees with the experimental residence time, providing a physical interpretation to our entirely data-driven protocol. In combination with appropriate enhanced sampling algorithms, this technique can be used for the initial exploration of ligand-dissociation pathways as well as for calculating path-specific thermodynamic and kinetic properties.


\end{abstract}

\section{Significance Statement}
Present-day computational drug design primarily relies on ligand-receptor binding free energies, despite the growing realization that drug residence time and unbinding pathways play key roles in determining in-vivo efficacy. We introduce an automated approach to classify ligand dissociation pathways using a powerful speech recognition algorithm called dynamic time warping. Combining enhanced sampling atomistic simulations with our path-classification algorithm, we distinguish various ligand unbinding pathways with $\sim$90\% accuracy and discover kinetically distinct dissociation channels that were indistinguishable through conventional analysis of the trajectories. We also predict exit-path specific ligand unbinding kinetics in quantitative agreement with experiments. Incorporating this information can transform rational drug design and help combat the emergence of drug-resistant mutations.

\section{Introduction}

Mechanistic understanding of protein-ligand interaction is of fundamental importance to the rational design of therapeutic drugs. Atomistic molecular dynamics (MD) simulations can provide insight into the mechanisms of protein-ligand association and dissociation, and are, therefore, used extensively in computer-aided drug design \cite{de2016role}. With the development of enhanced sampling algorithms \cite{rocchia2012enhanceddrugdesign} and coarse-grained models \cite{souza2020protein}, it is now possible to simulate physiologically relevant drug-receptor interactions at an affordable computational cost.

Most drug discovery efforts based on MD simulation have focused on calculating the binding free energy which indicates the thermodynamic stability of the ligand-receptor complex \cite{de2016role}. However, the kinetics and pathways of target recognition are critical determinants of the efficacy and selectivity of small-molecule drugs. It has been demonstrated, in a variety of targets, that drug residence time i.e. the ligand unbinding kinetics is better correlated to the in-vivo efficacy of pharmaceutical drug candidates \cite{Copeland2006Drug-targetOptimization,Copeland2015TheRetrospective,Guo2012FunctionalTime}. The significance of drug unbinding pathways has been realized in recent experimental and computational studies when Lyczek et al. and Shekhar et al. showed that a drug-resistant mutation in Abl kinase can reduce the activity of the anticancer drug Imatinib (Gleevec) by modifying only the drug unbinding pathway without changing its binding affinity \cite{lyczek2021mutation,shekhar2022protein}.


Despite the progress in developing physics-based algorithms for accurate prediction of free energy and kinetics of ligand-receptor binding, the study of molecular pathways is usually conducted at a qualitative level, through manual observation and interpretation of the MD trajectories. This can become impractical if screening of a large number of drug candidates for a given pharmaceutical target is required. Moreover, a large volume of MD simulation datasets are being made available recently for pharmaceutically relevant targets such as the proteins in the SARS-CoV-2 virus \cite{amaro2020community}. Significant human effort is required to analyze these datasets and understand the molecular processes involved. For example, manual analysis of multiple microseconds of MD trajectories has led to the discovery of the dominant pathways for the binding of beta-blocker drug molecules to the G-protein coupled receptors (GPCR) \cite{dror2011pathway}. Ansari et al. discovered, through visual inspection, two different pathways of benzamidine unbinding from trypsin protein \cite{Ansari2022WaterTrypsin}. The two paths differ only in the role of the hydration water molecules in facilitating the ligand release. However, the kinetics of these pathways differ by almost one order of magnitude.  

Only recently, efforts are being devoted to designing automated approaches to analyze the transition pathways sampled from MD simulation. Notable examples include the use of a t-distributed stochastic neighbor embedding (t-SNE) algorithm to project the drug unbinding trajectories in a low dimensional space followed by agglomerative clustering to classify the sampled pathways
\cite{Rydzewski2016MachineP450cam}. 
A variational autoencoder based latent-space path clustering (LPM) algorithm could also distinguish molecular transition pathways into parallel kinetic channels \cite{meng2017path}, making it easier to decipher molecular mechanisms \cite{Qiu2023efficient}. 
Wolf and coworkers have attempted not only to classify the ligand release pathways but also to compute pathway-specific free energy profiles and kinetics. To this effect, they used principal component analysis (PCA) and constructed a dendrogram of molecular trajectories \cite{bray2022ligand}. Pathway-dependent kinetics and free energy profiles are then computed using dissipation-corrected targeted molecular dynamics (dcTMD) \cite{wolf2018targeted,wolf2020multisecond} and a modified version of Jarzynski equality \cite{wolf2023path}. 
Deep learning techniques such as self-organizing maps (SOM) have also been tested for classifying ligand unbinding pathways. The resulting approach, known as PathDetect-SOM, could successfully distinguish different pathways of inhibitor dissociation from heat shock protein 90 (HSP90) \cite{Motta2022PathDetect-SOM:Simulations}. 

The currently available path classification algorithms have a few key limitations making them difficult to use in drug discovery. 
These methods often assume that different pathways are distinguishable when projected on a reduced dimensional space \cite{meng2017path,Qiu2023efficient,bray2022ligand,Rydzewski2016MachineP450cam}. However, during the process of binding or unbinding, a drug interacts with many residues of the target protein in a specific temporal order characteristic to the particular pathway. Therefore, it is not guaranteed that such complex pathways can be separated in a low-dimensional latent space.    
Moreover, considerable system-specific knowledge is necessary to successfully use these algorithms, making them unsuitable for large-scale applications. Often, it is also necessary to use Markov state modeling (MSM) and (or) transition path theory (TPT) to preprocess and coarse-grain the trajectory data before it can be used for classification, significantly increasing the complexity of the protocol.


In this work, we aim at addressing these challenges by constructing an automated data-driven algorithm for the classification of transition pathways for molecular systems.  Our approach is based on the powerful speech recognition algorithm, dynamic time warping (DTW) \cite{sakoe1978dynamic}, which also found applications in signal processing \cite{Muller2007}, gene sequence alignment \cite{skutkova2013classification}, handwriting recognition \cite{Tappert1990handwriting}, gesture recognition \cite{Kuzmanic2007gesture} and predicting trends in finance and econometrics \cite{mastroeni2021decoupling,orlando2022financial,orlando2022modelling}. DTW is suitable for comparing time series or sequences of different lengths and clustering them based on their degree of similarity \cite{aghabozorgi2015time}. 

At a fundamental level molecular reaction pathways are high dimensional time sequences of uneven length since, due to the inherent stochasticity, transition paths may take different times in different independent trajectories. Dynamic time-warping can take into account this issue and can reduce the manual effort of trajectory pre-processing to a large extent. This feature motivated us to use DTW to classify various reactive pathways directly from MD trajectories. This comes with a few notable advantages. First, no dimensionality reduction or MSM construction is necessary, making it possible to compare directly different trajectories. Moreover, the DTW classification algorithm is independent of the enhanced sampling algorithm used to explore the pathways. Therefore, by making a careful choice of the enhanced sampling algorithm, such as metadynamics or conformational flooding, one can recover pathway-specific properties (i.e. free energy or kinetics) without modifying the path-classification protocol. 

Using a combination of enhanced sampling and dynamic time warping, we show that different reactive pathways can be accurately distinguished for a model M{\"u}ller Brown potential, in the conformational transition of Alanine dipeptide, and for protein-ligand unbinding. We chose, as an example of ligand-receptor dissociation, the complex of benzene with the L99A mutant of T4 lysozyme, a system for which multiple ligand unbinding pathways have been reported in the literature \cite{nunes2018escape,capelli2019exhaustive,rydzewski2019finding,nunes2021ligand,mondal2018atomic,smith2018multi}. Therefore, it is possible to compare the performance of our automated path classification protocol against the analysis conducted by human experts. We could also calculate the unbinding timescale ($\tau_{\text{off}}$) for individual pathways, with the fastest dissociation timescales in quantitative agreement with the experimental residence time. Although designed to study ligand-receptor binding or unbinding, our protocol can be generalized to any molecular conformational transition.


\section{Dynamic Time Warping }
\begin{figure}[]
    \centering
    \includegraphics[width=0.8\textwidth]{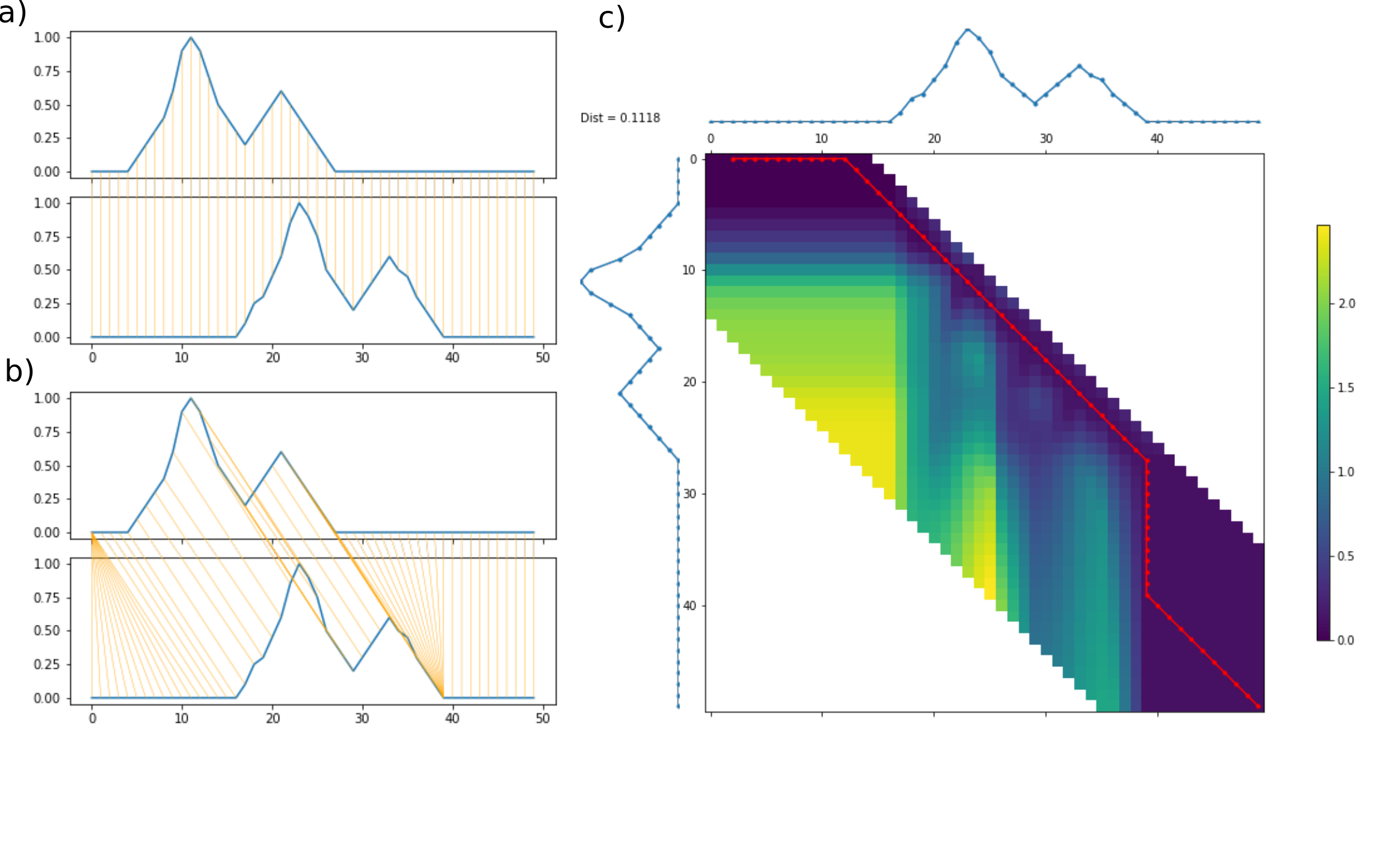}
    \caption{A schematic representation of dynamic time warping of two 1D time series. Panel (a) denotes the situation of a Euclidean warping where the signal at time = $t$ from time series 1 is compared directly with the signal at time $t$ of signal 2. (b) In dynamic time warping an optimal comparison between the two time-series is performed by minimizing the warping cost. This helps in comparing specific features of time series 1 with the corresponding signals in time series 2. In both (a) and (b) the x-axis indicates time and the y-axis indicates the signal of the time series. (c) The matrix $\mathbf{A}$ for comparing the two 1D time series. The colors indicate the value of the matrix elements. The minimum cost warping path is shown in red. In this example, the maximum deviation of the warping path from the diagonal is set to 15.    }
    \label{fig:schema}
\end{figure}

To classify molecular trajectories into different pathways it is necessary to construct a distance metric to measure the similarity between each pair of sampled transitions. 
Although root mean square deviation (RMSD) or similar metrics are widely used to compare static conformations of molecules, measuring the difference between dynamic trajectories can be challenging. The trajectories of protein-ligand dissociation can be represented as high-dimensional time series in molecular descriptor space composed of interatomic distances and contacts. The trajectories can have very different lengths and the key events in the unbinding mechanism (e.g. interaction between the ligand and protein residues) can take place at different times in different trajectories. Therefore a direct comparison of sampled conformations between different trajectories is not possible without the loss of the temporal information and compromising the accuracy of the classification algorithm (Fig. \ref{fig:schema}a). 

Dynamic Time Warping (DTW) is a data mining algorithm that is capable of comparing time series that are of unequal length (Fig. \ref{fig:schema}b). However, the utility of DTW in classifying molecular trajectories, has not been exploited and this algorithm is primarily used to distinguish one-dimensional time series. Generalizations of DTW to higher dimensions can be accomplished through the decoupling of each dimension of the time series (independent DTW or DTW$_{\text{I}}$). But, such approaches are not suitable for analyzing MD trajectories, as molecular degrees of freedom are often correlated to one another.

A recently proposed modification of DTW (or DTW$_{\text{D}}$, where the suffix D stands for dependent) can classify high-dimensional time series in correlated descriptor space \cite{shokoohi2017generalizing}. The DTW$_{\text{D}}$ method has been demonstrated to have a higher success rate compared to DTW$_{\text{I}}$ in gesture recognition from videos and for analyzing accelerometer data \cite{shokoohi2017generalizing}. Below we summarize the theoretical background of dynamic time warping, followed by a simple one-dimensional example. 

Let's assume there are two one-dimensional time series $Q = q_1,q_2,q_3,...,q_{m_1}$ and 
$C = c_1,c_2,c_3,...,c_{m_2}$ where in general $m_1 \neq m_2$. DTW creates a one-to-many alignment between these sequences by constructing a matrix $\mathbf{A} \in \mathbb{R}^{m1\times m2}$, whose elements are given by the squared Euclidean distance $d(q_i,c_j)$ between $q_i$ and $c_j$: $d(q_i,c_j) = (q_i - c_j)^2$. A warping path $P$ of length $T$ is defined as a contiguous set of matrix elements representing a mapping between time series $Q$ and $C$:
\[
P = p_1,p_2,p_3,...,p_T \;\;\; \text{such that} \;\; \text{max}(m_1,m_2) \leq T \leq m_1 + m_2 -1
\]

The warping path is subject to the following constraints: (1) $P$ must start and finish in diagonally opposite corners of the matrix $A$.
(2) The steps in the warping path are restricted to adjacent cells.
(3) The points in the warping path are monotonically ordered in time. In addition, a restriction may also be imposed on how far the path is allowed to deviate from the diagonal.

The cost $\mathcal{C}(P)$ associated with the warping path $P$ is defined as the square root of the sum of the matrix elements that are part of the warping path. 

\begin{equation}
    \mathcal{C}(P) = \sqrt{\sum_{t=1}^T p_t} = \sqrt{\sum_{A_{ij} \in P} A_{ij}}
\end{equation}

The DTW distance between two-time series is measured from the path that minimizes the warping cost:

\begin{equation}
 DTW(Q,C) = \text{min}_{P_k \in \mathbb{P}}\left\{\mathcal{C}(P_k)\right\}    
\end{equation}
where $\mathbb{P}$ is the set of all possible warping paths for the matrix $\mathbf{A}$.
The warping path with the least warping cost can be obtained using dynamics programming by evaluating the cumulative distance $D(i,j)$ given by: 

\begin{equation}
D(i,j) = d(q_i,c_j) + \text{min}\{D(i-1,j-1),D(i-1,j),D(i,j-1)\} 
\label{eq:dtw_dynamic}
\end{equation}
The objective of this exercise is to identify, between the two time-series, the best possible alignment which is represented by a valley of low $d(q_i,c_j)$ in the space of the matrix elements of $\mathbf{A}$ (Fig. \ref{fig:schema}c). After evaluating Eq. \ref{eq:dtw_dynamic} for all elements of matrix $\mathbf{A}$, the DTW distance between the series $C$ and $Q$ is computed as $DTW(Q,C) = D(m_1,m_2)$ i.e. the cumulative distance for the last element ($p_T$) of the matrix along both axes.

The dynamic time warping algorithm is demonstrated with a simple example in Fig. \ref{fig:schema}. We compare two 1D time series where the characteristic signal is virtually identical but out of phase. Through a direct comparison of the two signals at every time point, one obtains the Euclidean warping (Fig. \ref{fig:schema}a). This, however, fails to capture correctly the similarity between the two signals as the phase difference is not taken into account. DTW aims to circumvent this problem by finding an optimal alignment between the sequences to make a more meaningful comparison (Fig. \ref{fig:schema}b). To accomplish this, one constructs a matrix whose elements measure the pairwise Euclidean distance between every point in the two time-series. The dynamic programming algorithm (Eq. \ref{eq:dtw_dynamic}) then identifies a continuous path (colored red in Fig. \ref{fig:schema}c) of the smallest elements through the matrix. This process can be thought as analogous to identifying a 1D minimum energy manifold in a 2D potential energy surface. The DTW distance is then measured by summing all the elements of this optimal warping path.      

The DTW$_{\text{D}}$ generalizes this algorithm to multidimensional situation by modifying 
the distance measure $d(q_i,c_j)$ from Eq. \ref{eq:dtw_dynamic} as:
\begin{equation}
    d(q_i,c_j) = \sum_{m=1}^M (q_{i,m} - c_{j,m})^2
\end{equation}
Here $q_{i,m}$ is the $i$-th data point in the $m$-th dimension of an $M$ dimensional time series $Q$, and $c_{j,m}$ is the $j$-th data point in the $m$-th dimension of an $M$ dimensional time series $C$. To make the distance measure invariant of scale and offset, each dimension of the time series needs to be $z-$normalized. 

Based on the distances between trajectories computed from DTW$_{\text{D}}$ algorithm, a pairwise distance matrix is generated. K-medoid clustering is performed using this distance matrix to cluster the trajectories corresponding to different pathways. The choice of the clustering algorithm is inspired by the data mining and time series clustering literature \cite{hautamaki2008time,kaufman2009finding,liao2002understanding,liao2006adaptive}. K-medoid clustering identifies a physical trajectory (medoid) that is nearest to each cluster center, thereby increasing the interpretability of the overall protocol \cite{vuori2002comparison}. To identify the optimum number of clusters, we used the silhouette score metric \cite{rousseeuw1987silhouettes}. 



\section{Pathway Classification}

\begin{figure}
    \centering
    \includegraphics[width=\textwidth]{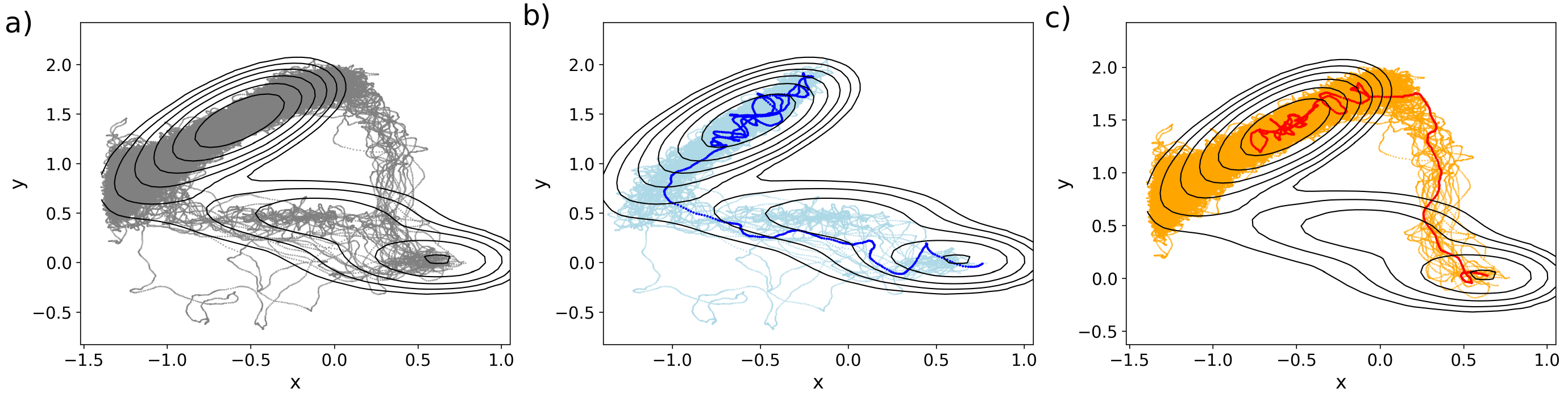}
    \caption{(a) All 40 transitions, sampled using OPES$_{\text{E}}$ simulation, for the 2D M{\"u}ller-Brown potential. Panel (b) and (c) represents the two clusters of pathways identified by the DTW$_{\text{D}}$ algorithm. The cluster medoid of each pathway is highlighted. }
    \label{fig:mueller}
\end{figure}
\noindent
\textbf{M{\"u}ller-Brown Potential:} To test the performance of DTW$_{\text{D}}$ in classifying MD trajectory data, we first applied it to the model system of M{\"u}ller-Brown potential. We could sample two different transition pathways using either the $x$ or the $y$ coordinate as CV in the exploratory variant of the On-the-fly Probability Enhanced Sampling \cite{Invernizzi2022ExplorationSampling}(OPES Explore or OPES$_{\text{E}}$). 
The trajectory data from two pathways are presented together to the DTW$_{\text{D}}$ algorithm, which could accurately distinguish the paths from the entire ensemble of trajectories represented as 2D time series $(x(t),y(t))$ (Fig. \ref{fig:mueller}). Despite the simplicity of the problem, this serves as a proof of concept of our protocol and sets the stage for its application to more complex systems.

\begin{figure}
    \centering
    \includegraphics[width=0.75\textwidth]{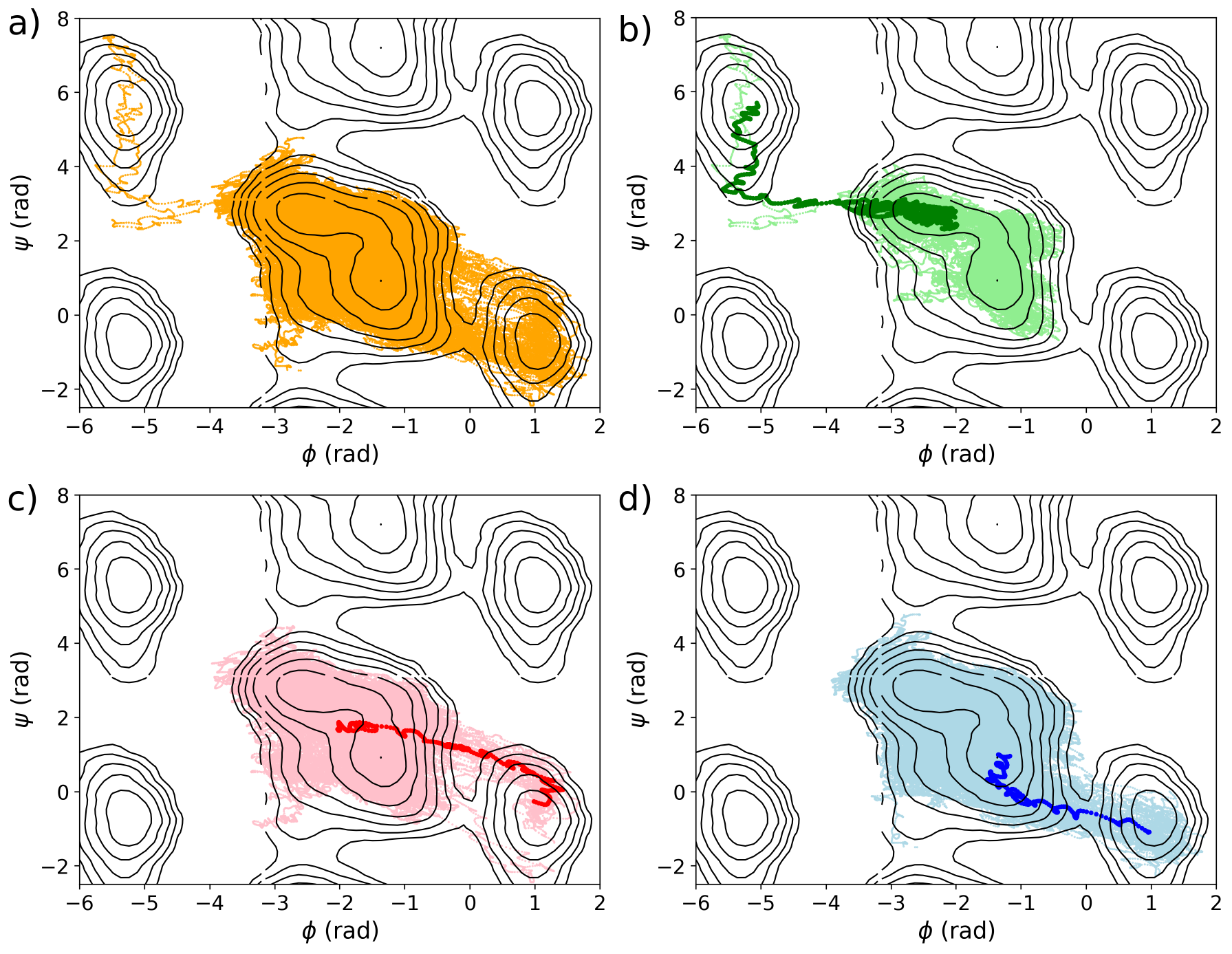}
    \caption{Panel (a) shows all 100 transitions from from $C7_{eq}$ state to the $C7_{ax}$ state of gas phase alanine dipeptide, sampled using OPES$_{\text{E}}$ algorithm. The trajectories are projected along the $\phi$ and $\psi$ torsion angles. The free energy surface for system is shown in black contour lines. Panel (b), (c), and (d) shows the trajectories belonging to the 3 path clusters identified by DTW$_{\text{D}}$ algorithm. The medoid of each cluster is highlighted in dark green, red, and dark blue colors respectively. The adjacent periodic images of the free energy contour is also shown for the ease of visualizing continuous pathways.}
    \label{fig:ala2}
\end{figure}
\noindent
\textbf{Alanine-dipeptide: }Next, we tested our approach to distinguish conformational transition paths in the gas phase alanine dipeptide. 
We sampled 100 transitions from the $C7_{eq}$ state to the $C7_{ax}$ state using OPES$_{\text{E}}$ algorithm with the $\phi$ torsion angle as CV. These transitions took place along 3 distinct pathways (Fig. \ref{fig:ala2}). Two of them proceed through the middle of the free energy surface (Fig. \ref{fig:ala2}c and d), while the third is a high energy path involving a reverse rotation around the $\phi$ torsion angle (Fig. \ref{fig:ala2}b). 
To avoid the inclusion of pre-existing system knowledge, we used a descriptor set consisting of 45 pair-wise distances between all non-bonded heavy atoms \cite{bonati2020data,bonati2021deep}. Even without any direct information about the $\phi$ and $\psi$ torsion angles that best describes the transitions, the DTW$_{\text{D}}$ algorithm successfully classified the pathways into three different clusters. In addition, the silhouette score metric indicated that the optimum number of clusters in this trajectory ensemble is 3. This result demonstrates that our protocol can correctly classify molecular paths using a generic descriptor set, and can automatically identify the number of pathways involved in the process. 

\begin{figure}
    \centering
    \includegraphics[width=0.4\textwidth]{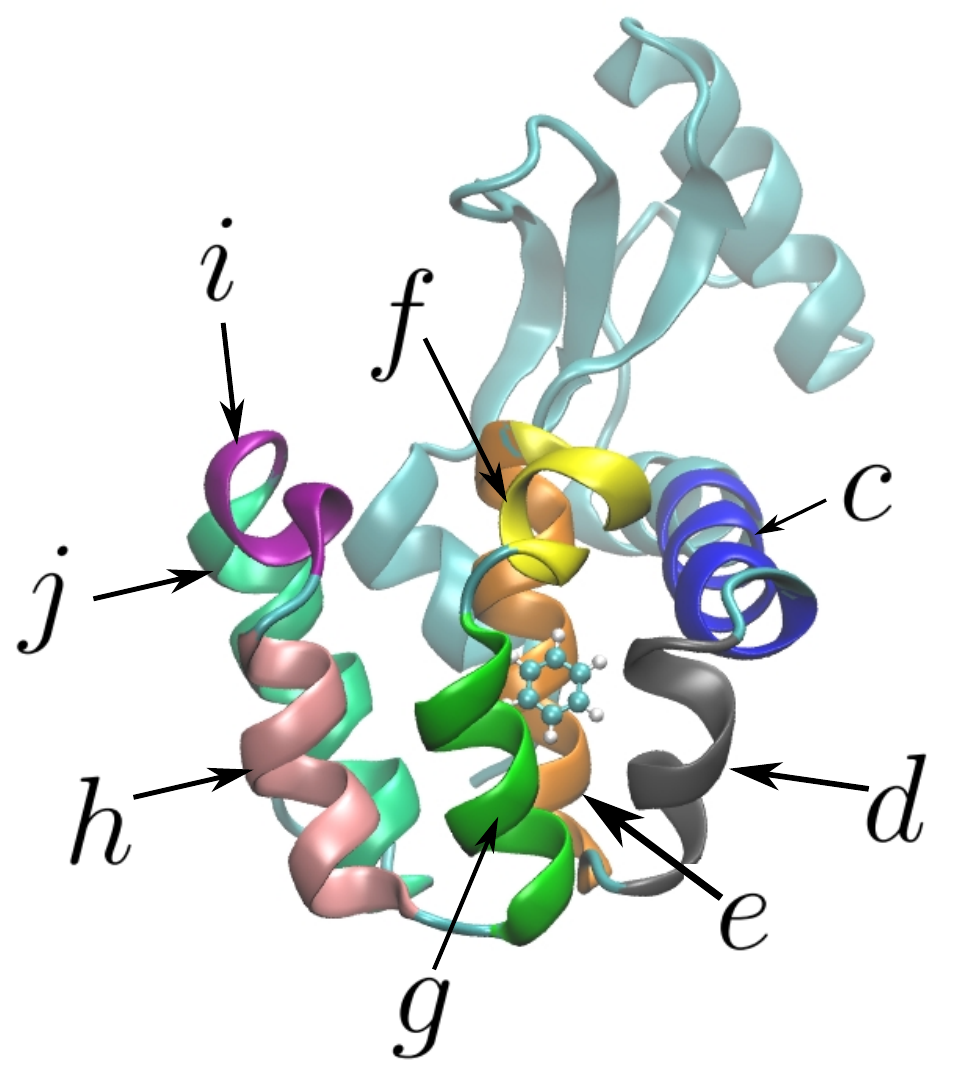}
    \caption{The prototypical protein-ligand complex of benzene bound to L99A mutant of T4 Lysozyme. The 8 $\alpha$-helices necessary to distinguish the various ligand release pathways are labeled $c$ through $j$. }
    \label{fig:t4l-helices}
\end{figure}

\begin{figure}
    \centering    \includegraphics[width=0.95\textwidth]{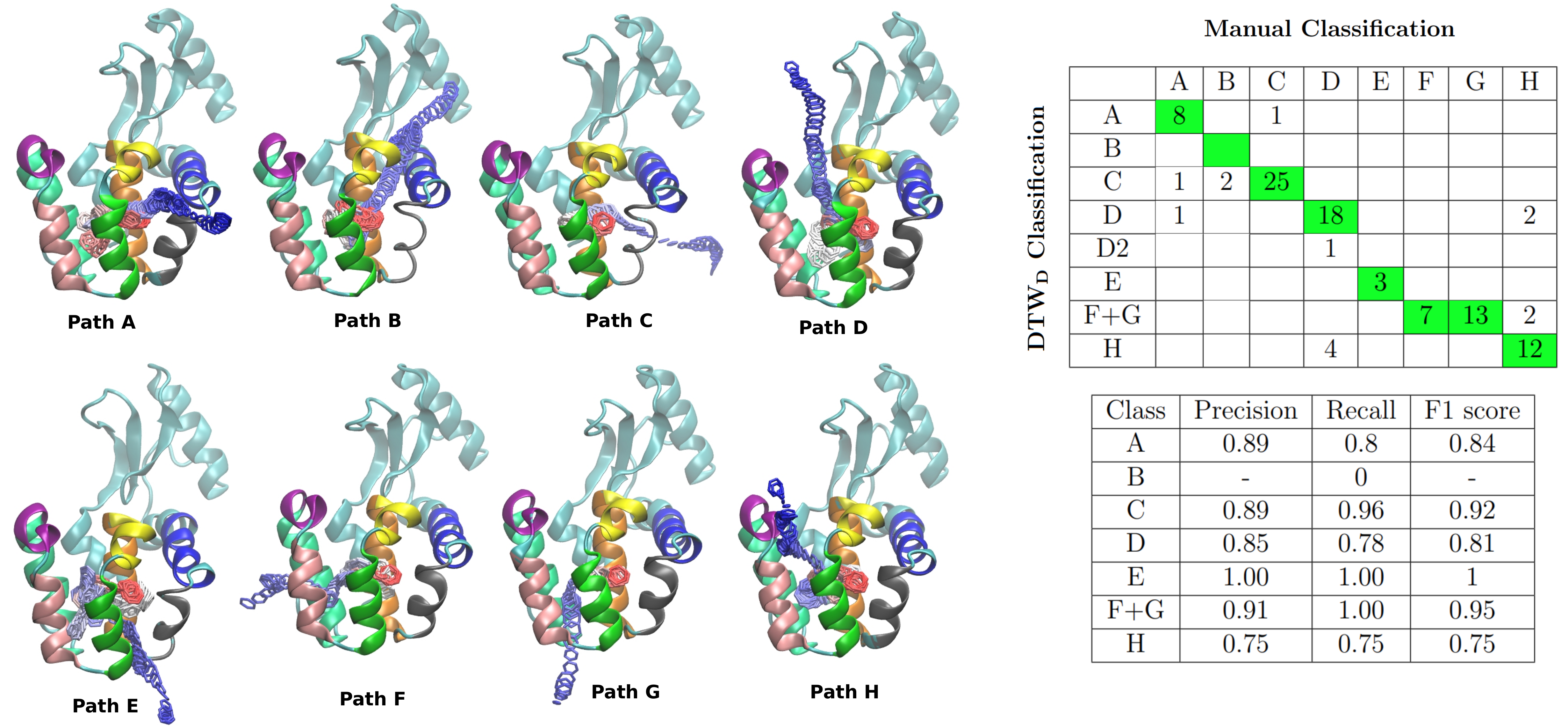}
    \caption{The 8 different pathways (A to H) of benzene unbinding from T4 lysozyme sampled using OPES$_{\text{E}}$ simulations. The benzene ring is colored according to the simulation time (red to blue in increasing order). The hydrogen atoms of benzene and the solvent molecules are omitted for clarity. The pathways shown in this figure are distinguished by visual inspection. In the right panel, we show the confusion matrix between the manual classification and DTW$_{\text{D}}$ classification. The true positive cells are colored green. Path F and Path G are considered together as they were not distinguished by DTW$_{\text{D}}$. }
    \label{fig:t4l-explore-paths}
\end{figure}
\noindent
\textbf{Protein-ligand Unbinding: }Now we investigated the accuracy of our approach in classifying unbinding pathways of benzene ligand from the L99A mutant of T4 Lysozyme. Using OPES$_{\text{E}}$ simulation, we could sample 100 dissociation events within a cumulative computational time of tens of nanoseconds. 
The ligand dissociation pathways can be manually classified into 8 pathways based on the interaction of the ligand with eight different helices of the protein (labeled $c$ to $j$ in Fig. \ref{fig:t4l-helices}). The transition paths sampled in the present study (Fig. \ref{fig:t4l-explore-paths}) are consistent with those of Capelli et al. \cite{capelli2019exhaustive}, Rydzensky and Valsson \cite{rydzewski2019finding}, and Nunes Alves and coworkers \cite{nunes2018escape,nunes2021ligand}. To provide a general description of the unbinding process, the coordination number of the ligand heavy atoms with the $C_{\alpha}$ atoms of each residue of the protein is included in the descriptor set resulting in a feature space with a dimensionality of 161. It should be noted here that the DTW algorithm has rarely been applied to such a high-dimensional time series clustering problem. However, our DTW$_{\text{D}}$ protocol could identify 7 clusters among the 100 ligand release trajectories, out of which 5 clusters directly correspond to 5 pathways (A, C, D, E, and H) identified through manual classification. The pathways F and G are combined within one cluster presumably because in these pathways benzene interacts with the $g$, $j$, and $h$ helices in a virtually identical manner except for changing course following the exit from the binding site. Instead, path B could not be identified as a separate cluster, likely due to its low population and its similarity with Path A and C in terms of the interaction of benzene with the protein residues in helix $c$ (Fig. \ref{fig:t4l-explore-paths}). To visualize the classification accuracy of the DTW$_{\text{D}}$ algorithm, we constructed a confusion matrix with the columns representing the pathways identified through manual classification and the rows representing those clustered by our automated protocol. The precision, recall, and F1 score values for every cluster are above 0.75 and an overall weighted F1 score of 0.86. This indicates a classification accuracy of 86\% which, considering the complexity of the protein-ligand dissociation mechanism, is noteworthy.

\section{Pathway dependent kinetics of ligand unbinding}
To calculate the pathway-specific kinetics of ligand dissociation, we sampled 137 benzene unbinding events using OPES flooding (OPES$_{\text{f}}$) \cite{Ray2022RareSampling} algorithm (See Methods and SI Appendix for details). The dissociation timescale ($\tau_{\text{off}}$) predicted from the entire set of trajectories (2.5 ms) is in good agreement with the experimental residence time of 1.1 ms. However, the $p$-value of the Kolmogorov-Smirnov (KS) test comparing the cumulative distribution of the ligand unbinding times with a perfect Poisson distribution is as low as 0.02. This indicates a deviation from a single exponential kinetic model, due to the different timescales involved in the exit of the ligand along different pathways.  

\begin{figure}
    \centering
    \includegraphics[width=0.9\textwidth]{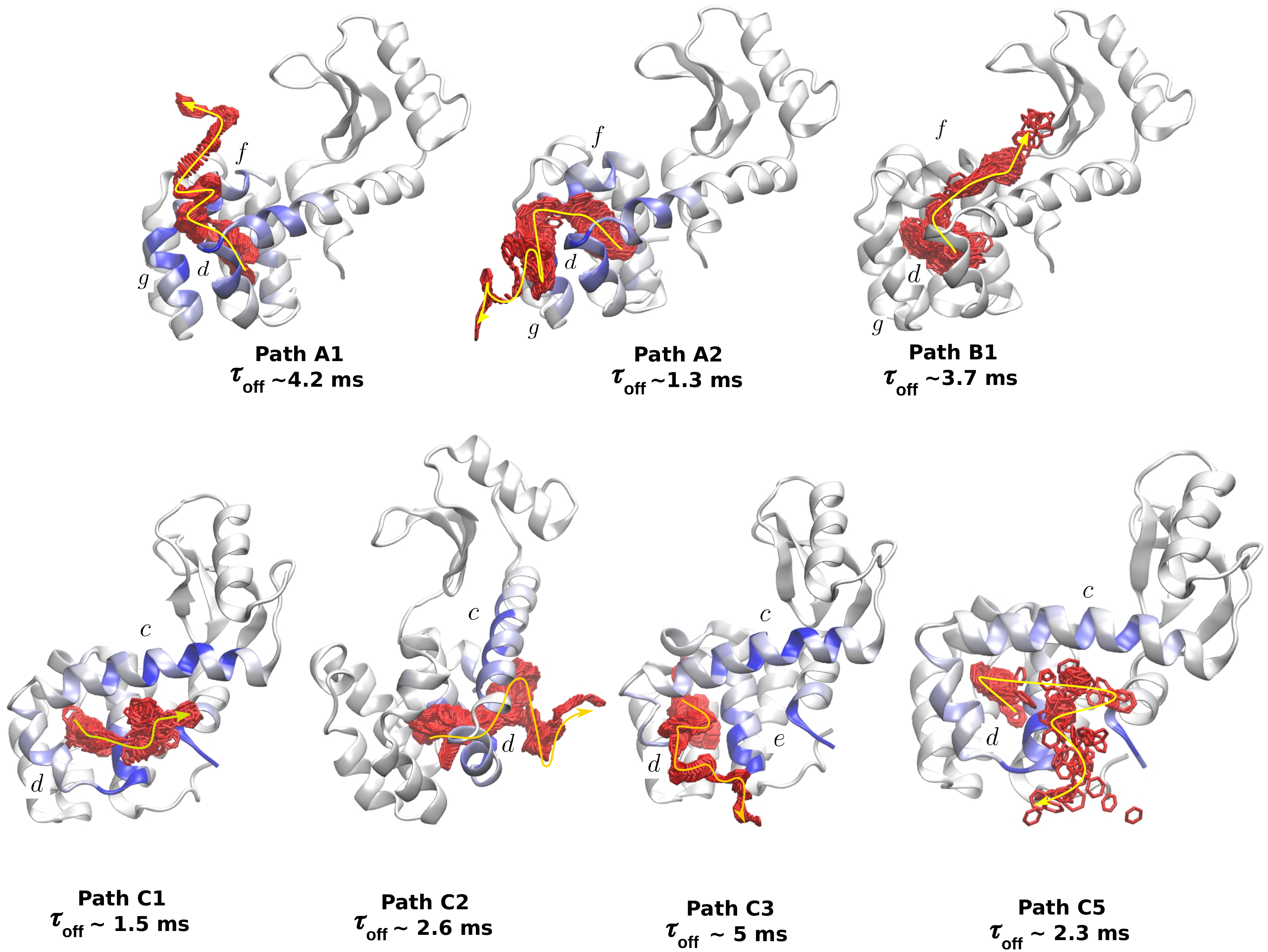}
    \caption{The medoids of the different dissociation channels within the Path A, B, and C, as identified by the DTW$_{\text{D}}$ analysis of the OPES$_{\text{f}}$ trajectories of the benzene-lysozyme complex. The unbinding timescales associated with each path cluster are denoted below each panel. The protein residues are colored according to their importance in distinguishing the path clusters within the same pathway (see text). Some of the helix labels are also included to aid visualization.   }
    \label{fig:t4l-path-kinetics}
\end{figure}

An identical path classification protocol based on the 161-dimensional molecular descriptor set could cluster the OPES$_{\text{f}}$ trajectories into 16 clusters, with one or more clusters assigned to each ligand release pathway. The classification accuracy, measured in terms of weighted F1 score is 96\%, an improvement over the results of the OPES$_{\text{E}}$ simulations. The increase in classification accuracy can be attributed to the fact that in OPES$_{\text{f}}$ trajectories spend more time in the transition region leading to better sampling of the exit paths. The pathways D, F, and H, for which there are only 1-3 samples, could also be classified correctly by our algorithm.

However, The most remarkable outcome of this work, is that we could calculate the exit-path-specific dissociation kinetics for this protein-ligand complex. The $\tau_{\text{off}}$ values differed significantly for different path clusters even within the same pathway. For example, the two clusters corresponding to Path A led to $\tau_{\text{off}}$ values that are different from each other by a factor of $\sim$3. Similarly, the unbinding timescales of four clusters for Path C range between $\sim$1.5 ms and $\sim$5 ms. This observation is noteworthy as these kinetically distinct path clusters are difficult to distinguish through manual visualization alone (Fig. \ref{fig:t4l-path-kinetics}).  

Most notably, the dissociation channels, identified through DTW$_{\text{D}}$ analysis, showed single exponential kinetics, demonstrated by KS test p-values significantly higher than 0.05 (between 0.3-0.99) (SI Fig. S6). Our classification protocol, therefore, distinguishes between kinetically distinct pathways without using any kinetic information in the training data. The fastest ligand dissociation path has a $\tau_{\text{off}}$ of 1.27 ms, which is in better agreement with the experimental unbinding rate constant ($k_{\text{off}}$ = 950 $\pm$ 20 s$^{-1}$), compared to the estimate obtained by taking together the dissociation timescales along different channels. This finding corroborates the pre-existing knowledge that the fastest unbinding pathway dominates the ligand residence time. 

We could also identify the role of individual residues that distinguish between these dissociation channels within individual unbinding pathways, as shown in SI Appendix.


\section{Discussions and Conclusions}

We demonstrate that a multidimensional dynamic time warping algorithm can accurately classify molecular transitions into different pathways in an automated data-driven manner. This protocol is suitable for the study of protein-ligand interaction where the kinetic properties are known to be dependent on the transition path. Our approach treats MD trajectories as time series in a high dimensional generic descriptor space, thereby reducing the manual effort and intuition involved in understanding complex ligand dissociation mechanisms. The DTW$_{\text{D}}$ protocol does not require any ad-hoc dimensionality reduction or the use of approximations like transition path theory, making it applicable to a wide range of molecular processes without system-specific knowledge.

To explore all possible pathways of ligand binding/unbinding, we use OPES$_{\text{E}}$ algorithm to sample transitions that are clustered afterwards based on the DTW$_{\text{D}}$ distance metric. This exercise provides only a qualitative description, but it can have important ramifications in computer-aided drug discovery. For example we utilized DTW to analyze the trajectories from different clusters to identify the key residues that differentiate the mechanism of ligand unbinding along different pathways. As shown by Shekhar et al. \cite{shekhar2022protein}, such information can provide an intuition about potential mutations that can develop resistance against potential therapeutic candidates. 

For a more quantitative insight, one can perform OPES$_{\text{f}}$ or OPES simulations, which, although computationally expensive, can compute the kinetics and free energy landscape, respectively, for individual pathways discovered by DTW$_{\text{D}}$. We demonstrated this by computing exit path specific $\tau_{\text{off}}$ for the benzene T4 lysozyme complex. The observed variance among the unbinding kinetics calls for a careful path classification analysis whenever attempting a ligand residence time calculation. Out of many possible transitions sampled from biased simulations, the kinetics along the fastest unbinding pathway is in excellent agreement with the experimental residence time. 

Ansari et al. have demonstrated that despite the apparent simplicity of the trypsin-benzamidine complex there are two different pathways of ligand unbinding with different dissociation timescales \cite{Ansari2022WaterTrypsin}. We see similar but more complex behavior in the case of benzene dissociation from T4 Lysozyme. The timescales for different pathways only vary by a factor of $\sim$3 indicating a difference in barrier heights by less than 1 kcal/mol, which is within the reputed uncertainties of the classical force fields. However, in situations where the difference between $\tau_{\text{off}}$ along different paths is higher, such an analysis can also provide a thermodynamic insight into the unbinding process. 

After an initial exploration it should also be possible to construct a path-specific collective variable \cite{branduardi2007b} to focus the sampling along specific pathways to obtain a more detailed understanding of the ligand unbinding mechanism. This is a potential avenue for future research. We also envisage the possibility of reducing the computational complexity of our path classification protocol using deep neural networks following the suggestions of Seshan \cite{seshan2022using}.

The present study on automated discovery and classification of ligand unbinding pathways will facilitate the incorporation of residence time and mechanistic descriptions of drug-target interactions within the realm of computer-aided drug discovery, a field that is heavily reliant on binding free energy calculations. This is an urgent necessity when limitations of free-energy-based screening are becoming increasingly apparent \cite{Copeland2006Drug-targetOptimization,Copeland2015TheRetrospective}. Our approach can facilitate a transition from the static free-energy-based screening of therapeutic candidates to the modeling of drug efficacy in the inherent out-of-equilibrium environment of the intracellular matrix. Moreover, the identification of protein residues that determine the ligand unbinding pathways, will help the pharmaceutical industry to adapt to the emergence of drug-resistant mutations \cite{shekhar2022protein,lyczek2021mutation}. Therefore, this work and follow-up studies in this area will be key steps toward potentially transforming the field of rational drug design. 





\section{Methods}
\subsection{M{\"u}ller Brown Potential}
Langevin dynamics simulations for the 2D M{\"u}ller Brown potential are performed using the \href{https://www.plumed.org/doc-master/user-doc/html/ves_md_linearexpansion.html}{\texttt{ves\_md\_linearexpansion}} module of PLUMED v2.9, with a setup identical to Ref. \citen{ray2023deep}. Twenty OPES$_{\text{E}}$ simulations, with $\Delta E = 20$ $k_BT$, were performed with either $x$ or $y$ as CV to sample transitions along the two pathways. Trajectories were stopped as soon as the system transited from the left to the right minima. Trajectory data from the transition region ($x$ $>$ -0.5 or $y$ $<$ 1.0) were analyzed using DTW$_{\text{D}}$. . 
\subsection{Alanine Dipeptide}
Gas phase Alanine Dipeptide is modeled using AMBER99SB-ILDN force field and simulated at 300 K using the GROMACS v2021.5 package patched with PLUMED v2.9. All simulation parameters are identical to Ref. \citen{Ray2022RareSampling}. OPES$_{\text{E}}$ simulations are performed along the $\phi$ torsion angle with a barrier parameter of $\Delta E = 40$ kJ/mol to sample 100 transition events. Trajectories were terminated upon reaching $C_{7ax}$ state from $C_{7eq}$ state. The trajectory data of the 45-dimensional descriptor space \cite{bonati2020data} from the transition region ($\phi > -1.0$ rad) is analyzed using DTW$_{\text{D}}$. 
\subsection{Benzene T4 Lysozyme Complex}

The topology and parameters for the benzene-bound L99A mutant of T4 Lysozyme were identical to Ref. \citen{capelli2019exhaustive} and are obtained from PLUMED NEST repository with \href{https://www.plumed-nest.org/eggs/19/017}{plumID:19.017}. The protein is modeled with the AMBER ff14SB force field and the ligand has been parametrized using the Generalized Amber Force field (GAFF) with RESP charges computed at the HF/6-31G(d) level of theory. All simulation conditions are identical to Ref. \citen{capelli2019exhaustive}. 

For the OPES$_{\text{E}}$ simulations, we used the same collective variables as Capelli et al. namely the spherical coordinates ($\rho$, $\theta$, $\phi$) of the center of mass of the ligand with respect to the center of mass of the binding pocket. The bias was deposited along all the three components of the CV space. The barrier parameter was set to $\Delta E$ = 50 kJ/mol. The simulations are terminated when the ligand reached $\rho$ $>$ 2.5 nm, which we designate as the unbound state.

For calculating kinetics, OPES$_{\text{f}}$  simulations (See details in SI Appendix) have been performed with a 2D CV space comprising of the spherical distance $\rho$ as well as the total number of contacts between the protein and the ligand defined as 
\begin{equation}
    c = \sum_{i \in A} \sum_{i \in B} \frac{1-(r_{ij}/d)^6}{1-(r_{ij}/d)^{12}}
    \label{eq:contact}
\end{equation}
Here A and B are the sets of non-hydrogen atoms of the protein and the ligand respectively and $d$ = 0.45 nm. The barrier parameter was set to $\Delta E$ = 30 kJ/mol and an excluded region is applied at $\rho$ $>$ 0.9 nm to avoid bias deposition in the transition state. To ensure the sampling of independent pathways, the simulation starting points were picked from a 20 ns unbiased simulation in the bound configuration. All enhanced sampling MD simulations were performed using GROMACS v2021.5 patched with PLUMED v2.9.

For each trajectory, the coordination number between the ligand heavy atoms and the C-$\alpha$ atoms of each of the 161 protein residues is stored at 1 ps interval. The contacts are evaluated using Eq. \ref{eq:contact} using $d$ = 0.8 nm. Only the portion of the trajectory with $\rho > 0.3$ nm and $c > 3$ was used for path classification to focus the emphasis on the transition region. The resulting 161-dimensional time-series were subsequently analyzed using DTW$_{\text{D}}$. 

It has been shown that at least 10 transitions are to be sampled to estimate the rates to gain statistical reliability \cite{shekhar2022protein}. Seven out of 16 clusters, identified by our approach, contained 10 or more unbinding events. This allowed us to calculate the ligand dissociation kinetics along these 7 channels (Fig. \ref{fig:t4l-path-kinetics}).

\subsection{Dynamic Time Warping}
The DTW$_{\text{D}}$ distance \cite{shokoohi2017generalizing} between all pair of trajectories were computed using the \texttt{dtw\_ndim} module of the python package \texttt{DTAIDistance}\cite{DTAIDistance}. The pairwise distance matrix is used to perform a k-medoids clustering of the trajectories using the FasterPAM \cite{schubert2021fast} algorithm employed in the \texttt{kmedoids} package \cite{schubert2022fast} in Python. Silhouette scores \cite{rousseeuw1987silhouettes} were also computed using the \texttt{kmedoids} code.




\begin{acknowledgement}
The authors thank Luigi Bonati, Pedro Juan Buigues Jorro, Sudip Das, and Valerio Rizzi for their comments and suggestions. The authors thank Riccardo Capelli for sharing the input files and the trajectory data from Ref. \citen{capelli2019exhaustive}. The authors declare no competing financial interest.

\end{acknowledgement}

\section{Data Availability Statement}
The simulation input files and analysis scripts are available from the GitHub repository: \url{https://github.com/dhimanray/DTW_Path_Classification.git}. They will also be uploaded to the PLUMED NEST repository. The raw data for dynamic time warping clustering and manual classification are provided in the supporting information.

\vspace{2cm}





\bibliography{references}

\end{document}